\documentclass[12pt]{article}
\usepackage{graphicx}
\usepackage{subfigure}
\usepackage{amsmath}
\usepackage{amssymb}

\begin{document}

\begin{center}
\begin{large}
\title\\{ \textbf{Isgur-Wise Function for Heavy Light Mesons in D dimensional Potential Model.}}\\\

\end{large}

\author\

\textbf{$Sabyasachi\;Roy^{\emph{1}}\footnotemark\;, B\: J\: Hazarika^{\emph{2}}\:and\:D\:K\:Choudhury^{\emph{1,2,3}}$ } \\\
\footnotetext{Corresponding author. On leave from Karimganj College,Assam, India. e-mail :  \emph{sroy.phys@gmail.com}}
\textbf{1}. Department of Physics, Gauhati University, Guwahati-781014, India.\\
\textbf{2}. Centre for Theoretical Studies, Pandu College, Guwahati-781012, India\\
\textbf{3}. Physics Academy of The North East, Guwahati-781014, India. \\

\begin{abstract}
We report results of a potential model for mesons in D space-time dimension developed by considering the quark-antiquark potential of Nambu-Goto strings. With this wave function, we have studied Isgur-Wise function for heavy-light mesons and its derivatives like slope and curvature. The dimensional dependence of our results and a comparative study with the results of 3+1 dimensional QCD are also reported.\\

Key words : Nambu-Goto potential, L\"{u}scher Term, space-time dimension. \\\
PACS Nos. : 12.39.-x , 12.39.Jh , 12.39.Pn.\\

\end{abstract}
\end{center}

\section{Introduction:}\rm
The two quark composite system (mesons),with one heavy quark, has been the focus of interest for many years [1]. In the semileptonic transitions for mesons[2], in the infinite quark mass limit, all the mesonic form factors can be expressed in terms of a single universal function, the Isgur-Wise(IW) Function [3].\\
The wave function for the heavy-light mesons has been calculated earlier within the framework of QCD potential model [4,5] with considerable accuracy. This has been deduced both with coulombic term in potential as parent and also with the linear confinement term as parent[6-9]. The characteristics like slope (charge radii ) and curvature ( convexity parameter ) of IW function has been reported in both the two cases with certain limitations. In this paper, we calculate the wave function of heavy-light mesons by solving D dimensional Schrodinger equation with potential inspired by Nambu-Goto strings.\\
Nambu-Goto action of bosonic string[10-12] predicts the quark-antiquark potential[13] to be:
\begin{equation}
V ( r ) = \sigma r +\mu_0 +\frac{\gamma}{r}
\end{equation}
The coefficient$ \gamma = - \frac{\pi(d-2)}{24} $  is the universal L\"{u}scher coefficient [14] of the L\"{u}scher term  $ \frac{\gamma}{r}$ , which depends upon the space-time dimension d. $\sigma$ is the string tension whose value is $ 0.178\; GeV^2$, $\mu_0$ is a regularisation dependent constant.\\
With this potential, considering L\"{u}scher term as parent, we employ higher dimensional Schrodinger equation, for getting the unperturbed wave function. Next, we consider linear confinement term as perturbation and find the total wave function for mesons using Dalgarno's method of perturbation.  This we take as a generalisation of earlier work [15] in 3-dimension, mathematically considering L\"{u}scher term in Nambu-Goto potential to be identical with coulombic term in QCD potential. Although linear term in this potential is the leading one,we justify our such consideration under the observation that the Luscher term becomes more and more significant as D goes on increasing. \\
With the wave function generated, we then deduce  Isgur-Wise function (IWF) and its derivatives (slope and curvature). As stated above, the same study has been carried out in three dimensional Euclidean space with Cornell potential[15]. In our work, we also make a comparative study of our results in higher dimension with those obtained with standard QCD potential. \\
The section 2 contains the essential formalism, section 3 , the results and the section 4 contains conclusion and remarks.

\section{Formalism:}

\subsection{Potential Model:}
For simplification of our work and for better comparison with Cornell potential, we put the Nambu-Goto potential ( equation (1)  )  as :
\begin{equation}
 V(r)=-\frac{\gamma}{r}+\sigma r +\mu_{0}
\end{equation}

\begin{center}
now with $\gamma = \frac{\pi(d-2)}{24}$.
\end{center}
Here d is the space-time dimension with $d=D+1$ , D being the space dimension.

We take $ -\frac{\gamma}{r}$ as parent and $\sigma r + \mu_0 $ as perturbation. Our unperturbed Hamiltonian [16] is then,
\begin{equation}
H_0=-\frac{\nabla_D^{2}}{2\mu}-\frac{\gamma}{r}
\end{equation}
with
\begin{equation}
H^{\prime}= \sigma r +\mu_0
\end{equation}
as perturbation. $\mu$ is the reduced mass of the meson.

\subsection{Wave function with L\"{u}scher term as parent:}
In D-dimension , the Schrodinger equation is [17-21]:
\begin{equation}
[-\frac{\hbar^{2}}{2\mu}\nabla_D^{2} +V_0(r)]\Psi(r,\Omega_D)=E\Psi(r,\Omega_D)
\end{equation}
with ,
\begin{equation}
 \nabla_D^{2}=\frac{1}{r^{D-1}}\frac{d}{dr}(r^{D-1}\frac{d}{dr})-\frac{\Lambda_D^{2}(\Omega_D)}{r^2}
 =\frac{d^{2}}{dr^{2}}+\frac{D-1}{r}\frac{d}{dr}-\frac{\Lambda_D^{2}(\Omega_D)}{r^{2}}
\end{equation}
and
\begin{equation}
\Psi(r,\Omega_D)=R(r)Y(\Omega_D)
\end{equation}
Here, $ \frac{\Lambda_D^{2}(\Omega_D)}{r^{2}}$ is a generalisation of the centrifugal barrier [22] for the case of D-dimensional space-time. \\
The eigen values of $ \Lambda_D^{2}(\Omega_D) $ are given by :
\begin{equation}
\Lambda_D^{2}(\Omega_D)Y(\Omega_D)=l(l+D-2)Y(\Omega_D)
\end{equation}
Here $Y(\Omega_D)$ and $R(r)$ are the spherical harmonics and radial wave function; $l$ is the angular momentum quantum number and E is the energy eigen value for the unperturbed wave function. \\
This gives the equation (4) in term of radial part as :
\begin{equation}
[\frac{d^{2}}{dr^{2}}+\frac{D-1}{2}\frac{d}{dr}-\frac{l(l+D-2)}{r^{2}}+\frac{2\mu}{\hbar^{2}}(E-V_0)]R(r)=0
\end{equation}

For $ l=0 $ and taking $\hbar=1 $, equation (9) yields,
\begin{center}
\begin{equation}
R^{\prime\prime}(r)+\frac{D-1}{r}R^{\prime}(r)+2\mu(E+\frac{\gamma}{r})R(r)=0
\end{equation}
\end{center}
Following reference [23] , we look for the solution of eqn (10) , by taking the radial wave function of the form :
\begin{equation}
R(r)=F(r)e^{-\mu \gamma r}
\end{equation}
Here, it is to be noted that the negative sign in the exponent ensures the square integrability at the origin and infinity. \\
Using equation (11) in (10), we obtain,
\begin{equation}
F^{\prime\prime}(r)+(\frac{D-1}{2}-2\mu\gamma)F^{\prime}(r)+(\mu^{2} \gamma^{2}-\frac{D-1}{2}\mu \gamma+2\mu E+\frac{2\mu\gamma}{r})F(r)=0
\end{equation}
Now, we take,
\begin{equation}
F(r)=\sum_{n=0}^{\infty}a_n r^{n}f(r,D)
\end{equation}
such that $f(r,D)=1$ at $D=3$. We take [ ] $ f(r,D)= r^{\frac{D-3}{2}}$, which satisfies this condition.
We find the radial solution as:
\begin{equation}
R(r)=\sum_{n=0} ^{\infty} a_n r^{n+\frac{D-3}{2}}e^{-\mu\gamma r}
\end{equation}
With this radial wave function we construct the wave function $ \Psi^{0}(r) $ for unperturbed Hamiltonian, with $ n=0 $ as :
\begin{equation}
\Psi^{0}(r)= N r^{\frac{D-3}{2}}e^{-\mu\gamma r}
\end{equation}
N is the normalisation constant, which we obtain from the condition:
\begin{equation}
\int_0^{\infty} 4\pi r^{2} \mid \Psi^{0} (r)\mid ^{2}dr =1
\end{equation}
Applying equation (15)in (16), we get,
\begin{equation}
N=[\frac{(2\mu\gamma)^{D}}{4\pi\Gamma (D)}]^{1/2}
\end{equation}
At $D=3$, this gives $ N= (\frac{\mu^{3}\gamma^{3}}{\pi})^{1/2} $ which is similar to the value obtained earlier  [15] for the corresponding case with QCD potential.
With this unperturbed wave function, we measure the eigen energy E as :
\begin{equation}
E=W^{0}=-\int_0^{\infty}\frac{\gamma}{r} 4\pi r^{2}\mid \Psi^{0}(r)\mid^{2}dr =-\frac{\mu\gamma^{2}}{D-1}
\end{equation}

\subsection{Wave function with linear term as perturbation:}
The first order perturbed eigen function $ \Psi^{\prime}(r)$ can be calculated using the relation [22] :
\begin{equation}
H_0 \Psi^{\prime}(r)+H^{\prime}\Psi^{0}(r)=W^{0}\Psi^{\prime}(r)+W^{\prime}\Psi^{0}(r)
\end{equation}
where,
\begin{equation}
W^{\prime}=\int_0^{\infty} 4\pi r^{2}H^{\prime}\mid \Psi^{0}(r)\mid^{2}dr =\frac{\sigma D}{2\mu\gamma}+\mu_0
\end{equation}
Then, from equation (19) we get,
\begin{center}
\begin{flushleft}
\begin{equation}
[-\frac{\hbar^{2}}{2\mu}\nabla_D^{2}-\frac{\gamma}{r}-W^{0}]\Psi^{\prime}(r) =(W^{\prime} -\sigma r -\mu_0)\Psi^{0}(r)
\end{equation}
\end{flushleft}
\end{center}

Taking $\hbar=1$ and expanding,
\begin{equation}
[\frac{d^{2}}{dr^{2}}+\frac{D-1}{2}\frac{d}{dr}+\frac{2\mu\gamma}{r}-2\mu W^{0}]\Psi^{\prime}(r)=2\mu (\sigma r+\mu_0 -W^{\prime})\Psi^{0}(r)
\end{equation}
From equation (22), following Dalgarno's method [25-27] of perturbation, we get [Appendix-A]:
\begin{equation}
\Psi^{\prime}(r)=  -\frac{\sigma D}{6\gamma}r^{2}r^{\frac{D-3}{2}}e^{-\mu\gamma r}
\end{equation}
With this perturbed wave function, we construct the total wave function as :
\begin{equation}
\Psi^{total}(r)=N_1[\Psi^{0}(r)+\Psi^{\prime}(r)] \\\
\end{equation}
Using equations (15) and (23)in (24):
\begin{equation}
\Psi^{total}(r)=N_1[1-\frac{\sigma D}{6\gamma}r^{2}]r^{\frac{D-3}{2}}e^{-\mu\gamma r}
\end{equation}

where $N_1$ is the normalisation constant for the total wave function and is obtained from :
\begin{equation}
\int_0^{\infty} 4\pi r^{2} \mid \Psi^{total} (r)\mid ^{2}dr =1  \\\
\end{equation}
as
\begin{equation}
N_1=\frac{1}{2\sqrt{\pi}[\frac{\Gamma(D)}{(2\mu\gamma)^{D}}-\frac{2\sigma D}{6\gamma}\frac{\Gamma(D+2)}{(2\mu\gamma)^{D+2}}+\frac{\sigma^{2}D^{2}}{36\gamma^{2}}\frac{\Gamma(D+4)}{(2\mu\gamma)^{D+4}}]^{1/2}}
\end{equation}
For $D=3$, $N_1$ gives the value of eqn. (12) of ref [15] when $\gamma$ is replaced by $\frac{4\alpha_s}{3}$ and $\sigma$ by $b$. \\
Now, if there is only coulombic term as parent, with no perturbation, then $ \sigma \rightarrow 0 $ . In this case eqn. (27) becomes:
\begin{equation}
N_1=\frac{1}{2\sqrt{\pi}[\frac{\Gamma(D)}{(2\mu\gamma)^{D}}]^{1/2}}= [\frac{(2\mu\gamma)^{D}}{4\pi\Gamma(D)}]^{1/2}
\end{equation}
This is our normalisation constant N for unperturbed wave function $ \Psi^{0}(r)$ as obtained in equation (17). \\
\subsection{Isgur-Wise Function:}
  Isgur-Wise function, $\xi( v,v^\prime )$ depends only upon the four velocities   $v_\nu $ and $v_{\nu^\prime} $  of heavy particle before and after decay. This $\xi( v,v^\prime )$  is normalized at zero recoil [29]. If y = $v_\nu $.$v_{\nu^\prime} $ , then, for zero recoil (y=1), $\xi(y)=1$. In explicit form IW function can be expressed as :
\begin{equation}
\xi(y)=1-\rho^2 (y-1) +C(y-1)^2
\end{equation}
$\rho^2$ is the slope parameter at y=1 given by -
\begin{equation}
\rho^2 = -\frac{\delta\xi (y)}{\delta y}|_{y=1}
\end{equation}
\begin{flushright}
$\rho$ is known as the charge radius. \\
\end{flushright}
C is the convexity parameter given by -
\begin{equation}
C=\frac{\delta^{2}\xi (y)}{\delta y^2}|_{y=1}
\end{equation}
The calculation of this IW function is non-perturbative in principle and is performed for different phenomenological wave functions of mesons [28]. This function depends upon the meson wave function and some kinematic factor, as given below :
\begin{equation}
\xi(y)=\int_0 ^\infty 4\pi r^2 |\Psi(r)|^2\cos(pr)dr
\end{equation}
where $\cos(pr)=1-\frac{p^2 r^2}{2}+\frac{p^4 r^4}{4}$ +$\cdot\cdot\cdot\cdot\cdot\cdot$  with $ p^2=2\mu^2 (y-1).$ Taking cos(pr) up to  $O(r^4)$ we get,

\begin{equation}
\xi(y)= \int_0 ^\infty 4\pi r^2 |\Psi(r)|^2dr - [4\pi\mu^2\int_0^\infty r^4|\Psi(r)|^2dr](y-1)+[\frac{2}{3}\pi\mu^4\int_0^\infty r^6|\Psi(r)|^2dr](y-1)^2
\end{equation}
Equations (28) and (32) give us :

\begin{eqnarray}
\rho^2 = [4\pi\mu^2\int_0^\infty r^4|\Psi(r)|^2dr] \\
C= [\frac{2}{3}\pi\mu^4\int_0^\infty r^6|\Psi(r)|^2dr] \\
\int_0 ^\infty 4\pi r^2 |\Psi(r)|^2dr =1
\end{eqnarray}
Equation (35) gives the normalization constants $ N \;\;$ and $\;\;  N^{\prime}$ for $\Psi^0 (r)$ and $\Psi^{total} (r)$ as obtained earlier in equations (17) and (27).
\subsection{Derivatives of Isgur-Wise function:}
Taking the unperturbed wave function $\Psi^{0}(r)$ from equation (17), we have calculated the slope and curvature of IW function $ \xi(y)$ as given below :
\begin{eqnarray}
\rho^2=\frac{D(D+1)}{4\gamma^{2}} \\\
C=\frac{D(D+1)(D+2)(D+3)}{96\gamma^{4}}
\end{eqnarray}
Thus, for $ D=3$, we get $ \rho^{2}=\frac{3}{\gamma^{2}} $ and $ C=\frac{15}{4\gamma^{4}}$ , which are the expressions (25) and (26) of ref. [15], if $\gamma$ and $\sigma$ are replaced by $\frac{4\alpha_s}{3}$ and $b$ of QCD potential as given below in eqn. (38).\\
\begin{equation}
V(r)= -\frac{4\alpha_s}{3r}+br+c
\end{equation}

Also,With total wave function $\Psi^{total}(r)$ we have calculated the slope and curvature.
\begin{flushleft}
\begin{eqnarray}
\rho^{2}=\mu^{2}\frac{[\frac{\Gamma(D+2)}{(2\mu\gamma)^{D+2}}-\frac{2\sigma D}{6\gamma}\frac{\Gamma(D+4)}{(2\mu\gamma)^{D+4}}+\frac{\sigma^{2}D^{2}}{36\gamma^{2}}\frac{\Gamma(D+6)}{(2\mu\gamma)^{D+6}}]}{[\frac{\Gamma(D)}{(2\mu\gamma)^{D}}-\frac{2\sigma D}{6\gamma}\frac{\Gamma(D+2)}{(2\mu\gamma)^{D+2}}+\frac{\sigma^{2}D^{2}}{36\gamma^{2}}\frac{\Gamma(D+4)}{(2\mu\gamma)^{D+4}}]} \\\
C=\frac{\mu^{4}}{6}\frac{[\frac{\Gamma(D+4)}{(2\mu\gamma)^{D+4}}-\frac{2\sigma D}{6\gamma}\frac{\Gamma(D+6)}{(2\mu\gamma)^{D+6}}+\frac{\sigma^{2}D^{2}}{36\gamma^{2}}\frac{\Gamma(D+8)}{(2\mu\gamma)^{D+8}}]}{[\frac{\Gamma(D)}{(2\mu\gamma)^{D}}-\frac{2\sigma D}{6\gamma}\frac{\Gamma(D+2)}{(2\mu\gamma)^{D+2}}+\frac{\sigma^{2}D^{2}}{36\gamma^{2}}\frac{\Gamma(D+4)}{(2\mu\gamma)^{D+4}}]}
\end{eqnarray}
\end{flushleft}
These at $ D=3 $ becomes:
\begin{eqnarray}
\rho^{2}=\mu^{2}\frac{[\frac{3.4}{(2\mu\gamma)^{2}}-\frac{\sigma }{\gamma}\frac{3.4.5.6}{(2\mu\gamma)^{4}}+\frac{\sigma^{2}}{4\gamma^{2}}\frac{3.4.5.6.7.8}{(2\mu\gamma)^{6}}]}{[1-\frac{\sigma }{\gamma}\frac{3.4}{(2\mu\gamma)^{2}}+\frac{\sigma^{2}}{4\gamma^{2}}\frac{3.4.5.6}{(2\mu\gamma)^{4}}]} \\\
C=\frac{\mu^{4}}{6}\frac{[\frac{3.4.5.6}{(2\mu\gamma)^{4}}-\frac{\sigma }{\gamma}\frac{3.4.5.6.7.8}{(2\mu\gamma)^{6}}+\frac{\sigma^{2}}{4\gamma^{2}}\frac{3.4.5.6.7.8.9.10}{(2\mu\gamma)^{8}}]}{[1-\frac{\sigma }{\gamma}\frac{3.4}{(2\mu\gamma)^{2}}+\frac{\sigma^{2}}{4\gamma^{2}}\frac{3.4.5.6}{(2\mu\gamma)^{4}}]}
\end{eqnarray}
$ \rho^{2}$ and $C$ expressions in equations (39) and (40) also give back equations (36) and (37) when $ \sigma\rightarrow 0 $. Again, at $D=3$, eqns (39) and (40) transform to eqns (36) and (38) of ref. [15] on replacing $\frac{4\alpha_s}{3}$ for $\gamma$ and $b$ for $\sigma$ , if we neglect the reletivistic effect of ref [15].

\section{Results:}
\begin{figure}[h]
    \centering
    \subfigure[$\xi(y)$ vs y for diff. D with $\Psi^{0}(r)$]
    {
        \includegraphics[width=3.0in]{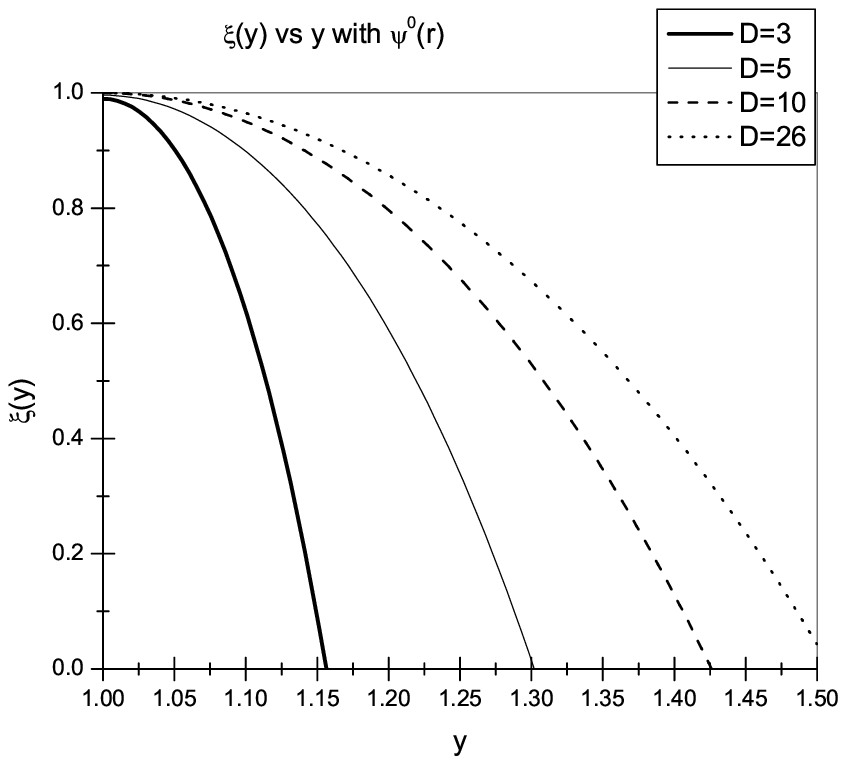}
        \label{fig:first_sub}
    }
        \subfigure[$\xi(y)$ vs y for diff. D with $\Psi^{total}(r)$ ]
    {
        \includegraphics[width=3.0in]{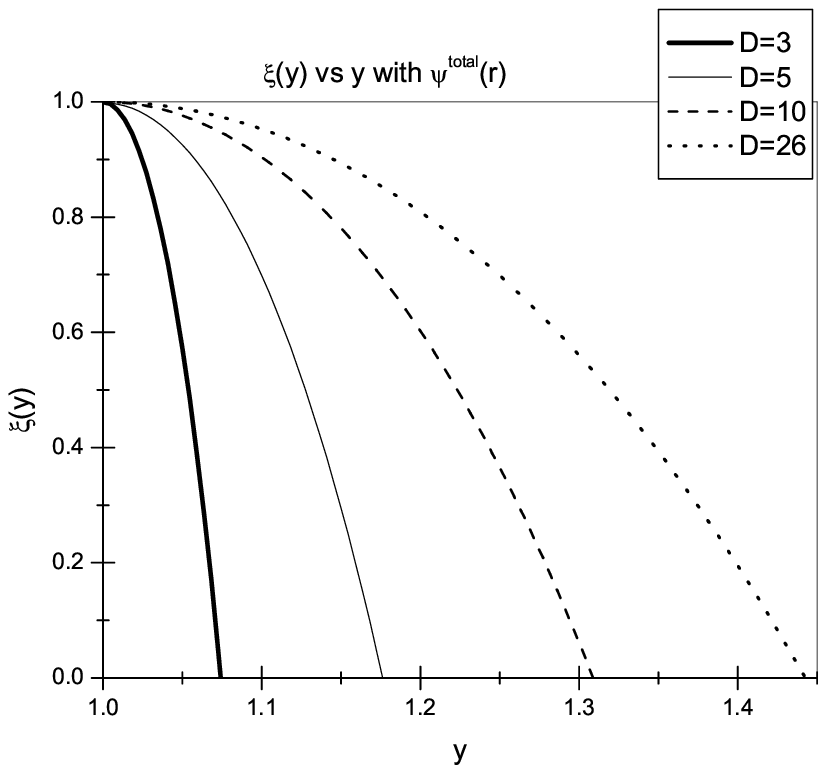}
        \label{fig:second_sub}
    }
\caption{Variation of IWF with y for B meson with diff. D values}
\label{fig:sample_subfigures}
\end{figure}
We have studied the variation of IW function with $ y $ for different space-time dimension D for B meson both for unperturbed and total wave functions taking string tension $\sigma=0.178 GeV^{2}$ (fig.1). The boundary condition $ \xi(1)=1 $ is found to be satisfied everywhere. \\
The variation of derivatives of IW function ( $ \rho^{2}$ and $C $ ) with space-time dimension D for B meson taking unperturbed wave function $ \Psi^{0}(r)$ and total wave function $ \Psi^{total}(r)$ are given in table 1.\\
\begin{table}[ht]
\begin{center}
\caption{Values of normalisation constant and derivatives of IW function for B meson.} \label{cross}
\begin{tabular}{|c|ccc|ccc|}
  \hline
 D &  &$ with  \Psi^{0}(r)$ &  &  &$ with  \Psi^{total}(r)$ &    \\

    & N & $\rho^2$ & C & $N_1$ & $\rho^2$ & C  \\
  \hline \hline
  3 & 0.0047 & 175.04 & 12766.00 & $2.49\times10^{-6}$ & 817.1 & 178802    \\
  4 & 0.0031 & 72.93 & 1861.70 & $6.5\times10^{-6}$ & 262.9 &   17584.2    \\
  5 & 0.00175 & 48.62 & 735.50 & $6.95\times10^{-6}$ & 146.18 &  5218.3   \\
  9 & 0.0001186 & 26.80 & 175.46 & $1.256\times10^{-6}$ & 54.44 &  650.2 \\
  10 & 0.00005856 & 25.07 & 148.56 & $6.94\times10^{-7}$ & 48.1 & 	498.7 \\
  26 &$5.13\times10^{-10}$ & 17.78 & 60.925 & $1.106\times10^{-11}$ & 23.67 & 	105.9 \\
  \hline

\end{tabular}
\end{center}
\end{table}
We find that with increase in D value, the slope and curvature decreases. With total wave function,the variation of  $\rho^2$ and $C$ with space-time dimension D for B meson is shown in fig.2.
\begin{figure}[h]
    \centering
    \subfigure[$\rho^{2}$ vs D with $\Psi^{total}(r)$]
    {
        \includegraphics[width=3.0in]{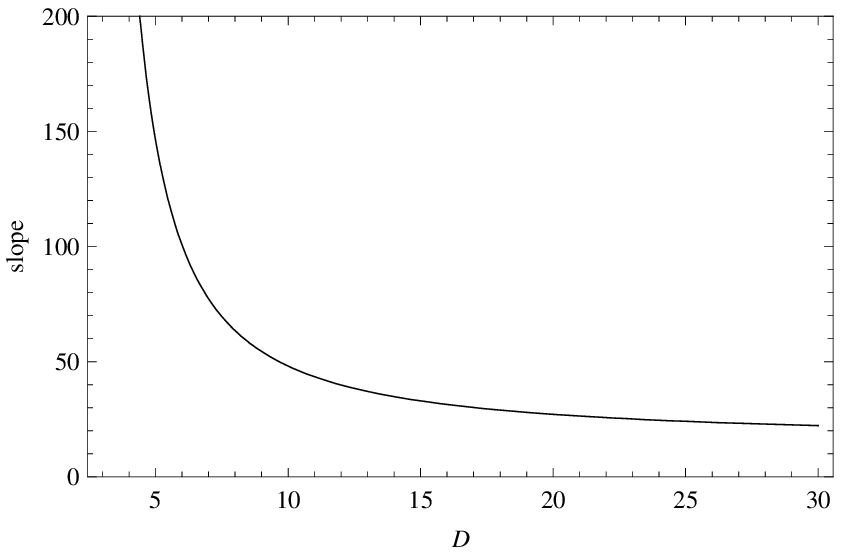}
        \label{fig:first_sub}
    }
        \subfigure[C vs D with $\Psi^{total}(r)$ ]
    {
        \includegraphics[width=3.0in]{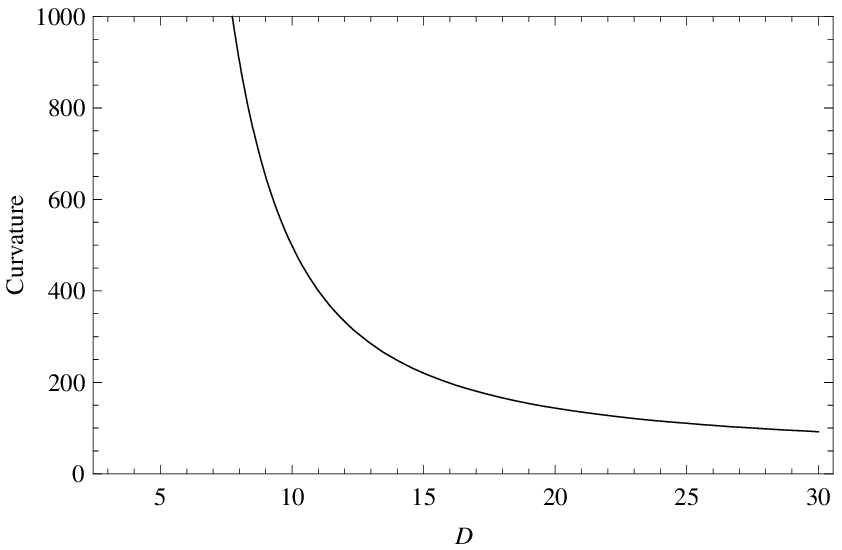}
        \label{fig:second_sub}
    }
\caption{Variation of $\rho^{2}$ and $C$ vs D for B meson }
\label{fig:sample_subfigures}
\end{figure}
As $D\rightarrow \infty$ we have also obtained the asymptotic forms of $ \rho^{2}$ and $C$ from equations (40) and (41).
\begin{eqnarray}
\rho^{2}_{asym}=\mu^{2}\frac{\frac{1}{(2\mu\frac{\pi}{24})^2}-\frac{2\sigma}{6(\frac{\pi}{24})}\frac{1}{(2\mu\frac{\pi}{24})^4}+\frac{\sigma^2}{36(\frac{\pi}{24})^2}\frac{1}{(2\mu\frac{\pi}{24})^6}}
{1-\frac{2\sigma}{6(\frac{\pi}{24})}\frac{1}{(2\mu\frac{\pi}{24})^2}+\frac{\sigma^2}{36(\frac{\pi}{24})^2}\frac{1}{(2\mu\frac{\pi}{24})^4}}\\
C_{asym}=\frac{\mu^{4}}{6}\frac{\frac{1}{(2\mu\frac{\pi}{24})^4}-\frac{2\sigma}{6(\frac{\pi}{24})}\frac{1}{(2\mu\frac{\pi}{24})^6}+\frac{\sigma^2}{36(\frac{\pi}{24})^2}\frac{1}{(2\mu\frac{\pi}{24})^8}}
{1-\frac{2\sigma}{6(\frac{\pi}{24})}\frac{1}{(2\mu\frac{\pi}{24})^2}+\frac{\sigma^2}{36(\frac{\pi}{24})^2}\frac{1}{(2\mu\frac{\pi}{24})^4}}
\end{eqnarray}
The asymptotic values of $ \rho^{2}$ and $C$ are found to be 14.5865 and 35.4608 respectively, for B meson.

\section{Conclusion and remarks:}
From table 1 and fig. 2, we find that, with increasing D , $ \rho^{2}$ and $C$ values go on decreasing eventually reaching the asymptotic limit. We have shown our result for B meson only as a representative case. Similar pattern of result is being found to follow for other B and D sector heavy-light mesons like $ D, D_s, B_s, B_c$. \\  Although expressions for $\rho^{2}$ and $C$ give back the corresponding expressions obtained with standard QCD potential for $D=3$ (ref.[15]) by replacing $\gamma$ with $\frac{4\alpha_s}{3}$ and $\sigma$ with $b$, but the values of $ \rho^{2}$ and $C$ is higher in the present case for $D=3$ . This is due to much lower value of $\gamma$ for lower D , as compared to that of the corresponding term $\frac{4\alpha_s}{3}$ in the previous work with QCD potential (ref.[15])\\
However, at higher D when $\gamma$ becomes more and more dominant, our $ \rho^{2}$ and $C$ values go on decreasing. This higher dimensional approach also in turn supports our consideration of treating Luscher term as parent with confinement term as perturbation in the present formalism.\\
Lastly, we conclude by making the optimistic comment that the picture will possibly improve at lower values of D, if we treat linear potential term as parent with Luscher term as perturbation.Development of formalism with the latter approach is under consideration.

\paragraph{Acknowledgement :\\ }
\begin{flushleft}
SR acknowledges the support of University Grants Commission, Govt. of India in terms of fellowship under FDP scheme to pursue his research work at Gauhati University. Authors extend heartiest thanks to the authority of Pandu College, Guwahati for providing necessary facilities.
\end{flushleft}

\appendix

\numberwithin{equation}{section}
\begin{center}
\section{Appendix}
\end{center}

From equation (22), the Schrodinger equation for first order perturbation is:
\begin{equation}
[\frac{d^{2}}{dr^{2}}+\frac{D-1}{2}\frac{d}{dr}+\frac{2\mu\gamma}{r}-2\mu W^{0}]\Psi^{\prime}(r)=2\mu (\sigma r+\mu_0 -W^{\prime})\Psi^{0}(r)
\end{equation}
Following Dalgarno's method [25-27] of perturbation, we obtain :
\begin{equation}
\Psi^{\prime}(r)=(\sigma r + \mu_0)R(r)
\end{equation}
Taking,
\begin{equation}
R(r)=G(r)e^{-\mu\gamma r}
\end{equation}
and then, expanding $ G(r)$ in terms of power series,
\begin{equation}
G(r)=\sum_{n=0}^{\infty} A_n r^{n+\frac{D-3}{2}}
\end{equation}
we ultimately obtain:
\begin{equation}
\Psi^{\prime}(r)= (\sigma r+\mu_0)\sum_{n=0}^{\infty} A_n r^{n+\frac{D-3}{2}}e^{-\mu\gamma r}
\end{equation}
This upon expansion gives-
\begin{eqnarray}
\Psi^{\prime}(r)=[\mu_0 A_0r^{0}+(\mu_0A_1+\sigma A_0)r^{1}+(\mu_0A_2+\sigma A_1)r^{2}+  \nonumber \\
(\mu_0A_3+\sigma A_2)r^{3}+(\mu_0A_4+\sigma A_3)r^{4}+.....]r^{\frac{D-3}{2}}e^{-\mu\gamma r} \\
\Psi^{\prime}(r)=\sum_{n=0}^{\infty}C_n r^{n} r^{\frac{D-3}{2}}e^{-\mu\gamma r}
\end{eqnarray}
where,
\begin{eqnarray}
C_0=\mu_0 A_0\;\;\;and\\
C_{l+1}=\mu_0 A_{l+1} + \sigma A_l
\end{eqnarray}
with $l=0,1,2,3,....$. \\
Using (A.7), (A.8) and (A.9) in (A.1) , we get,

\begin{eqnarray}
C_{l+1}=0 \; for\;\; l\neq 1 \;\;and\\
C_2=\frac{\mu}{3}(\mu_0-W^{\prime})
\end{eqnarray}

The equation (A.7) then reduces to:
\begin{equation}
\Psi^{\prime}(r)= [\mu_0 A_0 +\frac{\mu}{3}(\mu_0-W^{\prime})r^{2}]r^{\frac{D-3}{2}}e^{-\mu\gamma r}
\end{equation}
Now, using equation (20), viz, $\mu_0-W^{\prime}= -\frac{\sigma D}{2\mu\gamma}$, equation (A.12) becomes:

\begin{equation}
\Psi^{\prime}(r)= [\mu_0 A_0 -\frac{\mu}{3}\frac{\sigma D}{2\mu\gamma}r^{2}]r^{\frac{D-3}{2}}e^{-\mu\gamma r}
\end{equation}
As $A_0$ is undetermined, we set $\mu_0 A_0$ to be zero. This results in :
\begin{equation}
\Psi^{\prime}(r)=-\frac{\sigma D}{6\gamma}r^{2}r^{\frac{D-3}{2}}e^{-\mu\gamma r}
\end{equation}

\end{document}